\documentclass[3p,times,procedia]{elsarticle}
\usepackage{nupha_ecrc}

\volume{00}

\firstpage{1}

\journalname{Nuclear Physics A}

\runauth{Mace, Mueller, Schlichting, Sharma}

\jid{nupha}

\jnltitlelogo{Nuclear Physics A}

\usepackage{amssymb}
\usepackage{graphicx}  % needed for figures
\usepackage{dcolumn}   % needed for some tables
\usepackage{bm}        % for math
\usepackage{amsmath}
\usepackage{comment}
\usepackage[hidelinks]{hyperref}
\usepackage{graphicx}
\usepackage{subfig}
\usepackage{color}
\usepackage{slashed}
\usepackage{url}
\usepackage{hyperref}
\usepackage{lipsum}
\usepackage[numbers]{natbib}

\definecolor{darkturq}{RGB}{0, 206,209}

\newcommand{\Fig}[1]{Fig. \ref{#1}}

\newcommand{\xs}{\mathbf{x}}
\newcommand{\ib}{\mathbf{i}}
\newcommand{\xt}{\mathbf{x}_{\bot}}
\newcommand{\rsph}{r_{\text{sph}}}
\newcommand{\tsph}{t_{\text{sph}}}
\usepackage[figuresright]{rotating}

\begin{document}

\begin{frontmatter}



%% Instructions from Editor: Please use the following \dochead only in the preprint version (e-print arXiv etc.); 
%% use empty \dochead{} when submitting to Nuclear Physics A!
\dochead{XXVIth International Conference on Ultrarelativistic Nucleus-Nucleus Collisions\\ (Quark Matter 2017)}
%% e.g. \dochead{17th International Conference on Dynamical Processes in Excited States of Solids}

\title{Simulating chiral magnetic effect and anomalous transport phenomena in the pre-equilibrium stages of heavy-ion collisions}


\author[MM1,MM2]{Mark Mace}
\author[NM]{Niklas Mueller}
\author[SoS]{S\"{o}ren Schlichting}
\author[MM2]{Sayantan Sharma}
%\runauth[MM1,MM2]{Mark Mace}
%\runauth[NM]{Niklas Mueller}
%\runauth[SoS]{S\"{o}ren Schlichting}
%\runauth[MM2]{Sayantan Sharma}
%\

\address[MM1]{Physics and Astronomy Department, Stony Brook University, Stony Brook, NY 11974, USA}
\address[MM2]{Physics Department, Brookhaven National Laboratory, Bldg. 510A, Upton, NY 11973, USA}
\address[NM]{Institut f\"{u}r Theoretische Physik, Universit\"{a}t Heidelberg, Philosophenweg 16, 69120 Heidelberg, Germany}
\address[SoS]{Department of Physics, University of Washington, Seattle, WA 98195-1560, USA}

%\author{Mark Mace}
%\email{mark.mace@stonybrook.edu}
%\affiliation{Physics and Astronomy Department, Stony Brook University, Stony Brook, NY 11973, USA}
%\affiliation{Physics Department, Brookhaven National Laboratory, Bldg. 510A, Upton, NY 11973, USA}
%\author{Niklas Mueller}
%\email{n.mueller@thphys.uni-heidelberg.de}
%\affiliation{Institut f\"{u}r Theoretische Physik, Universit\"{a}t Heidelberg, Philosophenweg 16, 69120 Heidelberg, Germany}
%\author{S\"{o}ren Schlichting}
%\email{sslng@uw.edu}
%\affiliation{Department of Physics, University of Washington, Seattle, WA 98195-1560, USA}
%\author{Sayantan Sharma}
%\email{sayantans@quark.phy.bnl.gov}
%\affiliation{Physics Department, Brookhaven National Laboratory, Bldg. 510A, Upton, NY 11973, USA}

\begin{abstract}
We present a first principles approach to study the Chiral Magnetic Effect during the pre-equilibrium stage of a heavy-ion collision. We discuss the dynamics of the Chiral Magnetic Effect and Chiral Magnetic Wave based on real-time lattice simulations with dynamical (Wilson and Overlap) fermions simultaneously coupled to color and electromagnetic fields. While for light quarks we observe a dissipation-less transport of charges as in anomalous hydrodynamics, we demonstrate that for heavier quarks the effects of explicit chiral symmetry breaking 
lead to a significant reduction of the associated currents.
\end{abstract}


\end{frontmatter}

% \linenumbers

\section{Introduction}\label{sec:intro}
\begin{figure}
\centering
  \includegraphics[width=.75\textwidth]{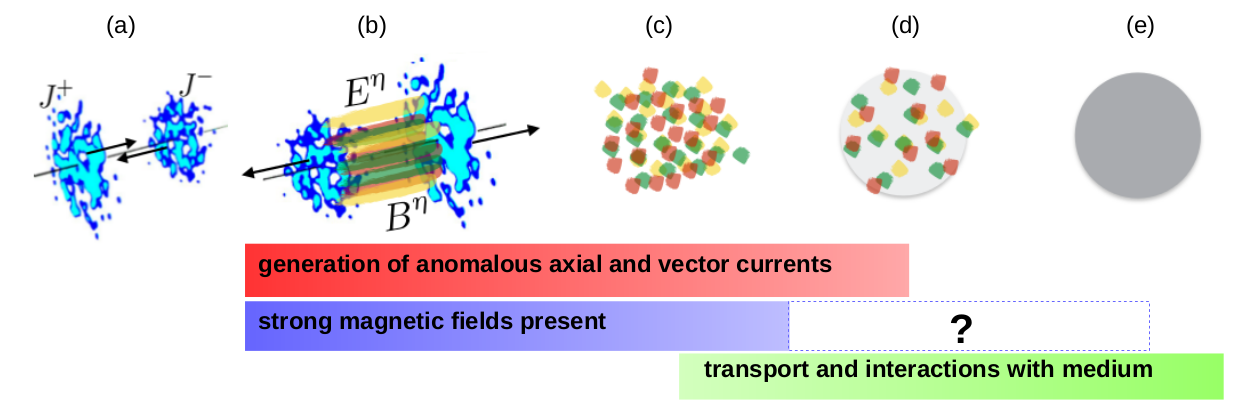}%
\caption{Schematic overview of the space-time dynamics of anomalous transport phenomena in heavy-ion collisions. Strong magnetic fields induced by the spectators are present during the pre-equilibrium stage, where anomalous currents are first created. Based on the weak coupling description of high-energy collisions in the Color Glass Condensate framework~\cite{McLerran:1993ni} the early time dynamics (a-c), can be described by classical-statistical simulations~\cite{Berges:2013eia,Gelis:2013rba,Kurkela:2012hp}, employed to study anomalous transport phenomena in this work. Subsequently, as the system expands, it transitions to the stage % where it is dilute and weakly coupled 
(d), which can be described using an effective kinetic description~{\cite{Baier:2000sb,Kurkela:2015qoa}}, with recent extensions to include anomalous transport processes via chiral kinetic theory~\cite{Son:2012wh,Stephanov:2012ki,Mueller:2017lzw}. Ultimately, the system approaches local thermal equilibrium (e) and can be described using anomalous hydrodynamics~\cite{Son:2009tf,Hirono:2014oda,Jiang:2016wve}.
}
\label{fig:cmeoverview}
\end{figure}

Novel transport phenomena associated with the interplay of quantum anomalies and magnetic fields and vorticity have created excitement across the physics community~\cite{Kharzeev:2007jp,Fukushima:2008xe,Kharzeev:2015znc}. Experimental searches for such anomalous transport phenomena in heavy-ion collisions have revealed intriguing hints at possible signals, however the situation remains unclear because of potentially large background effects~\cite{Abelev:2009ac,Khachatryan:2016got,Chatterjee:2014sea}.  Despite the fact that various theoretical techniques~\cite{Tanji:2016dka,Son:2009tf,Hirono:2014oda,Jiang:2016wve,Son:2012wh,Stephanov:2012ki,Mueller:2017lzw} have been developed to study anomalous transport processes, a robust theoretical prediction of the signal remains challenging. One of the key difficulties in this regard is due to the short lifetime ($\sim$1 fm/c) of the magnetic field in off-central heavy-ion collisions~\cite{Skokov:2009qp,McLerran:2013hla}, forcing anomalous transport processes to take place predominantly during the pre-equilibrium phase. Based on the recent progress in understanding the early-time dynamics of high-energy heavy-ion collisions~\cite{Gelis:2010nm}, a compact summary of the space-time evolution in the pre-equilibrium phase is given in \Fig{fig:cmeoverview}. While a combination of different theoretical techniques is required to describe the entire evolution, the early-time dynamics (a-c) can be accurately described using a classical-statistical field theory description, which can be solved using first principles lattice techniques.

In this proceeding, we report first steps towards a quantitative description of anomalous transport phenomena during the non-equilibrium stage from classical-statistical real-time lattice simulations. Based on a brief introduction to the formalism in Sec. 2, we highlight the key results of our present studies \cite{Mueller:2016ven,Mace:2016shq} in Sec. 3 and conclude with an outlook to future applications in heavy-ion collisions and towards a quantitative understanding of anomalous transport phenomena in high-energy heavy-ion collisions.

\section{Non-equilibrium lattice simulations}\label{sec:lattice}
Based on a classical-statistical description of the bosonic gauge-field degrees of freedom, we study the non-equilibrium production of fermions without any further approximations. Most importantly, the fermion degrees of freedom are treated fully quantum mechanically by discretizing the theory on a lattice and numerically solving the operator Dirac equation~\cite{Aarts:1999zn,Kasper:2014uaa},
\begin{eqnarray}
\label{eqn:driaceqn}
i \gamma^{0} \partial_t \hat{\psi}_x=(-i\slashed{D}^{s}+m)\hat{\psi}_x\;.
\end{eqnarray}
Here $\slashed{D}^{s}$ is either formulated in terms of Wilson-Dirac (see also~\cite{Aarts:1999zn,Kasper:2014uaa,Mueller:2016aao} for related studies) or Overlap fermion (see \cite{Mace:2016shq}) discretization schemes. In the case of Wilson fermions, we employ tree-level operator improvements to explicitly cancel lattice artifacts of order {$\mathcal{O}(a^{2n-1})$ (in practice we use a scheme accurate to order $\mathcal{O}(a^3)$)}, where $a$ is the lattice spacing
\begin{eqnarray}
-i\slashed{D}^{s}_{W}\hat{\psi}_\xs=\frac{1}{2} \sum_{n,i} C_{n}
\Big[ \Big(-i\gamma^{i}- n r_{w} \Big) U_{\xs,+ni} \hat{\psi}_{\xs+n\ib}
+2 n r_{w} \hat{\psi}_{\xs} - 
\Big(-i \gamma^{i}+n r_{w} \Big)U_{\xs,-ni} 
\hat{\psi}_{\xs-n\ib} \Big]\;.  
\end{eqnarray}
The operator valued fermion field, $\hat{\psi}$, is decomposed into a complete set of operators acting on the initial state, each multiplying a time dependent complex valued wavefunction, making the system of equations tractable on a computer. We then solve the time evolution of each wave function in the background of a single topological transition~\cite{Moore:2010jd,Mace:2016svc}, that we construct explicitly. Details of the implementation, such as the definitions of observables are described in~\cite{Mueller:2016ven,Mace:2016shq}.
\section{Non-equilibrium dynamics of anomalous effects}\label{sec:results}
\begin{figure}
\centering
\includegraphics[width=.5\textwidth]{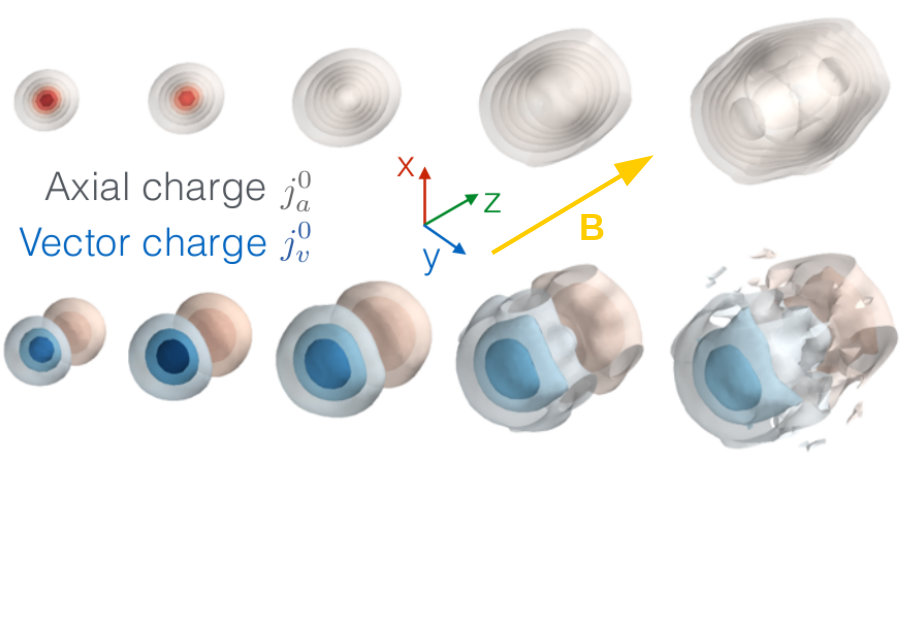}
\hspace{0.03\textwidth}
\includegraphics[width=.45\linewidth]{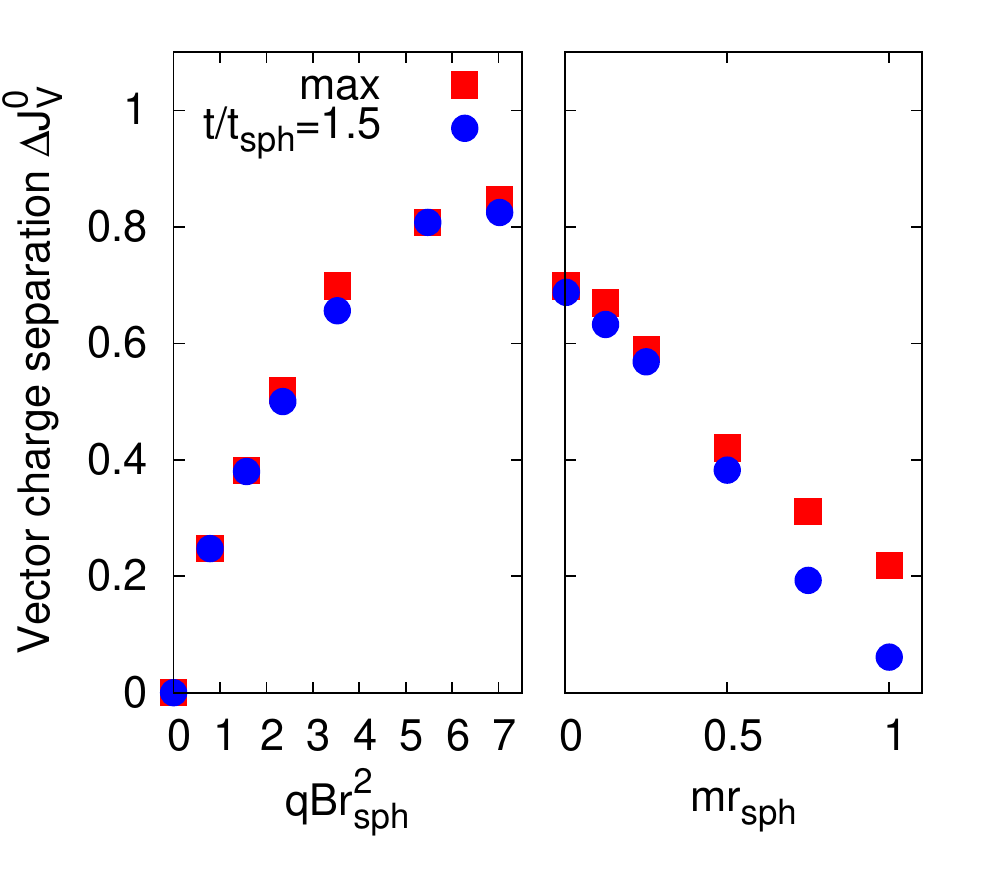}
\caption{Left: Profiles of the axial and vector densities for light quarks at different times of the real-time evolution for
fermions with $m \rsph=1.9 \cdot 10^{-2}$ for strong magnetic field $q B \rsph^{-2}=7.0$ at times $t/\tsph=0.6,0.9,1.1,1.3$. Center and right: Magnetic field and mass dependence of vector charge separation. The typical inverse size of the sphaleron is $r_{\text{sph}}^{-1}=$ 200-500 MeV~\cite{Mueller:2016ven,Mace:2016shq}.}
\label{fig:profilesmasses}%
\end{figure}

Based on the setup described in Sec.~\ref{sec:lattice}, we will now discuss the dynamics of axial charge generation during a sphaleron transition and the subsequent transport of axial and vector charge densities via the Chiral Magnetic Effect (CME) and Chiral Separation Effect (CSE). Our results are compactly summarized in the left panel of \Fig{fig:profilesmasses}, where we plot profiles of the axial and vector charges during and after a sphaleron transition in the presence of a strong magnetic field and very light quarks. 
By means of the axial anomaly, $\partial_\mu j^\mu_a=-(g^2/8\pi^2)\,\text{tr}F_{\mu\nu}\tilde{F}^{\mu\nu}$, 
the sphaleron transition leads to production of axial charge concentrated at the position of the sphaleron. In the presence of a magnetic field, the CME generates a vector current, resulting in a dipole like separation of vector charges along the direction of the magnetic field. As the system evolves, the vector charge imbalance at the edges of the dipole in turn creates an axial current due to the CSE. Ultimately, the simultaneous excitation of CME and CSE leads to the formation of a Chiral Magnetic Wave (CMW) {\cite{Burnier:2011bf}}, continuing to transport vector and axial charges along the magnetic field direction.
%
%Initially, a domain of axial charge is anomalously produced from the sphaleron transition. The CME then generates a vector %current, resulting in the depicted vector charge dipole. As the system evolves, the CME current transports the dipole apart, %creating a vector charge imbalance, and an axial current through the CSE, where ultimately an axial charge dipole is also %created. This simultaneous excitation of CME and CSE is known as the Chiral Magnetic Wave (CMW). %The separation of electric %charge along the magnetic field is observed and the emergence of chiral density waves is seen due to the interplay of the CME %and CSE~\cite{Mueller:2016ven,Mace:2016shq}.

In the center panel of \Fig{fig:profilesmasses} the magnetic field dependence of the vector charge separation $\Delta J_v^0=\int_{z\ge 0}dz ~d^2 \xt\, j_v^0(t,x)$ is shown, and as predicted by the CME, charge separation rises linearly for small magnetic field. 
At large magnetic fields the behavior is asymptotic, as the
vector charge separations saturates to unity. 

Since a finite quark mass leads to an explicit breaking of chiral symmetry, the axial anomaly equation is modified for massive quarks to include an explicit violation term, resulting in dissipation of axial charge. Indeed, we find that the explicit chiral symmetry breaking drastically reduces the amount of axial charge produced during the sphaleron transition. As the CME is approximately proportional to the axial charge imbalance, explicit chiral symmetry breaking results in a significant reduction of the vector charge separation observed in the right panel of \Fig{fig:profilesmasses}. Drastic differences between heavy and light fermions emerge already for modest fermion masses and anomalous transport essentially ceases to exist for $m r_{\text{sph}} \gtrsim 0.75$ ($r_{\text{sph}}$ is the size of the sphaleron), i.e. as soon as the quark mass becomes comparable to the other relevant scales in the problem. Even though such effects appear to be irrelevant for the two light flavors over the lifetime of a heavy-ion collision, our results suggest that strange quarks should not contribute significantly to anomalous transport processes.

Our microscopic simulations enable an explicit study of the constitutive relations for currents in anomalous hydrodynamics,
$j_{v,a}^\mu=n_{v,a}u^\mu + \sigma^B_{v,a} B^\mu$, by extracting the ratios between vector (axial) densities and axial (vector) currents at sufficiently late times. Interestingly we find that these ratios are not time-independent constants and away from the strong field limit differ significantly from unity ~\cite{Mueller:2016ven,Mace:2016shq}. Most importantly, we observe that the CME and CSE currents are not generated instantaneously. Given that the lifetime of the magnetic field in heavy ion collisions is short, this finite relaxation time must be taken into account for macroscopic descriptions of anomalous transport phenomena.

\section{Conclusions and Outlook}\label{sec:conclusion}
We studied the non-equilibrium dynamics of axial and vector charges and currents in the presence of non-Abelian and Abelian fields using real time lattice simulations. Our microscopic approach has enabled us to study anomalous transport from first principles. In comparison to analytic results for strong magnetic fields~\cite{Kharzeev:2007jp}, we find significant alterations at weak and moderate magnetic field strength, where  ratios between vector (axial) densities and axial (vector) currents clearly differ from unity, indicating only a partial alignment of spins with the magnetic field. Our findings also suggest that the onset of the CME and CSE is not instantaneous, and a finite relaxation time for the generation of the CME and CSE should be taken into account. Similarly, we find that for massive fermions significant dissipation effects exists, which effectively prohibit anomalous transport via the CME and CSE.
%
%We studied the non-equilibrium dynamics of axial and vector charges and currents in the presence of non-Abelian and Abelian %fields using real time lattice simulations. This microscopic approach has enabled us to study anomalous transport coefficients %for hydrodynamic simulations and we find significant alterations at weak and moderate magnetic fields. Our findings suggest that %a finite relaxation time for the generation of the CME and CSE must be taken into account. Moreover, we find that the ratios %between
%vector (axial) densities and axial (vector) currents clearly differ from unity, indicating only partial polarization of the %medium. Additionally, we find that for massive fermions significant dissipation effects exists, contributing and effectively %prohibiting the otherwise topologically protected transport via the CME and CSE. 

Since the lifetime of the magnetic field in a heavy ion collision is expected to be very short, it is essential to study the early time dynamics of the aforementioned phenomena. Our findings and technical progress in terms of novel algorithmic techniques form the basis for future studies towards a quantitative understanding of the CME at early times. %Our simulations allow us to address the two of the biggest theoretical uncertainties for the CME signal, as they enable us to determine the axial and vector charge and current distributions at earliest times, crucial inputs for hydrodynamic simulations. 
Besides the direct importance to early-time dynamics, the results of these microscopic studies can interface with macroscopic descriptions of anomalous transport, e.g. chiral kinetic theory or anomalous hydrodynamics. These macroscopic studies should include as an initial condition pre-equilibrium axial and vector charges and currents, which can be provided from real-time lattice simulations. Moreover, we can study the electromagnetic response of the dynamically created medium at earliest time and thus investigate the lifetime of the magnetic field by including fermionic back-coupling~\cite{Mueller:2016aao}, a crucial input for phenomenology.

 % \mfm{we already say this once}Additionally, based on our current simulations, dissipative effects due to finite quark mass as well as finite relaxation times for the generation of CME and CSE currents should be taken into account in any realistic description and we expect to obtain further insights in this regard from future studies.

\textit{Acknowledgements:} We are supported in part by the U.S. Department of Energy under Grant No. DE-SC0012704 (M.M., Sa.S.), DE-FG88-ER40388 (M.M.), DE-FG02-97ER41014 (So.S.), and by the Studienstiftung des Deutschen Volkes, by the DFG Collaborative Research Centre SFB 1225 (ISOQUANT) (N.M.), and within the BEST Topical Collaboration (M.M.). This research used resources of the NERSC, a DOE Office of Science User Facility supported by the Office of Science of the U.S. Department of Energy under Contract No. DE-AC02-05CH11231. Part of this work was performed on the Steinbruch Center ForHLR funded by
the Ministry of Science, Research and the Arts Baden-W\"urttemberg and the Deutsche Forschungsgemeinschaft and the USQCD clusters at Fermilab.



%% References


\bibliographystyle{elsarticle-num}



\end{document}